\documentclass[aps,pra,twocolumn,floatfix,superscriptaddress]{revtex4-1}

\usepackage[utf8]{inputenc}
\usepackage[T1]{fontenc}
\usepackage{float} 
\usepackage{color}  

\usepackage{graphicx}
\usepackage{amsmath}
\usepackage{amssymb}
\usepackage{bbm}
\usepackage{indentfirst}
\usepackage{dcolumn}
\usepackage{soul}
\usepackage{mathrsfs}
\usepackage{xcolor}
\usepackage{bigints}

\begin{document}

\title{Reduced thermal stability of antiferromagnetic nanostructures}

\author{Levente R\'{o}zsa}
\email{rozsa.levente@physnet.uni-hamburg.de}
\affiliation{Department of Physics, University of Hamburg, D-20355, Hamburg, Germany}
\author{Severin Selzer}
\affiliation{Department of Physics, University of Konstanz, D-78457 Konstanz, Germany}
\author{Tobias Birk} 
\affiliation{Department of Physics, University of Konstanz, D-78457 Konstanz, Germany}\author{Unai Atxitia}
\email{unai.atxitia@fu-berlin.de}
\affiliation{Department of Physics, University of Konstanz, D-78457 Konstanz, Germany}
\affiliation{Freie Universit\"{a}t Berlin, Fachbereich Physik, Arnimallee 14, D-14195 Berlin, Germany}
\author{Ulrich Nowak}
\affiliation{Department of Physics, University of Konstanz, D-78457 Konstanz, Germany}
\date{\today}

\begin{abstract}
Antiferromagnetic materials hold promising prospects in novel types of spintronics applications. Assessing the stability of antiferromagnetic nanostructures against thermal excitations is a crucial aspect of designing devices with a high information density. Here we use theoretical calculations and numerical simulations to determine the mean switching time of antiferromagnetic nanoparticles in the superparamagnetic limit. It is demonstrated that the thermal stability is drastically reduced compared to ferromagnetic particles in the limit of low Gilbert damping, attributed to the exchange enhancement of the attempt frequencies. It is discussed how the system parameters have to be engineered in order to optimize the switching rates in antiferromagnetic nanoparticles.
\end{abstract}

\maketitle

\section{Introduction}

In the field of spintronics, the storage, transfer and processing of information is based on the spin magnetic moment of electrons. Conventional spintronic devices are mainly based on ferromagnetic (FM) systems. However, recent advances in understanding and controlling antiferromagnetic (AFM) materials have led to an increasing interest in AFM spintronics \cite{ParkNatMaterials11,JungwirthNatNano16,WadleyScience16,OlejnikScienceAdvance18,NatPhyReview,GomonayReview,Balz}. Possible advantages of spintronic devices based on AFM materials include their lack of stray fields, which normally destroys single-domain states and leads to an interaction between bit patterns; the low susceptibility to external fields; and the rich choice of new materials, including a variety of AFM insulators. Moreover, AFM spin dynamics are found to be faster than those of FMs \cite{NelePRL2017,Selzer,GomonayPRL16,OlejnikScienceAdvance18}. 

For many applications, the size of magnetic structures will have to be scaled down to the nanometer regime, where, eventually, thermal excitations will reduce the stability of the magnetic state. In single-domain FM nanoparticles this is known as the superparamagnetic limit \cite{Bean}, where the whole structure can be described as a single macroscopic magnetic dipole. Besides their technological relevance, superparamagnetic particles have found their uses in biomedical applications \cite{Pankhurst} as well as in rock magnetism \cite{Rancourt}. Analogously, a single-domain AFM nanoparticle may be described by a macroscopic N\'{e}el vector, being the difference of the two sublattice magnetizations. The spontaneous switching of the N\'{e}el vector under thermal fluctuations constitutes the superparamagnetic limit in AFMs. In this context, it was shown recently \cite{Meinert} that thermally activated superparamagnetic reversal enhances the current-induced switching rates in AFM Hall cross devices. Furthermore, AFM nanoparticles play an important role in biological molecules such as the natural \cite{Allen} and synthetic forms \cite{Meldrum} of the iron-storage protein ferritin, and in the field of geochemistry \cite{Jambor}.

The thermal stability of FM nanoparticles has been studied extensively in the past \cite{WernsdorferPRL1997,BodePRL2004,GaraninPRB2008,Kleibert2014,Kleibert2017,Skumryev2003}. An analytical formula for the thermal switching rate in the superparamagnetic limit was first given by Brown \cite{Brown} based on the stochastic Landau--Lifshitz--Gilbert equation \cite{Landau1935,Gilbert2004}. 
The mean switching time in AFM nanoparticles has been investigated in significantly less works so far \cite{LothScience2012}, and the analytical studies \cite{Raikher2008,Ouari2010} have been restricted to the case of uncompensated AFMs with a finite magnetization. For current technological applications of compensated AFMs, a simple but accurate formula explaining the role of the interaction parameters in the reversal process seems to be lacking.

Here we theoretically investigate the switching rate in compensated AFM nanoparticles.
By deriving an analytical expression, it is demonstrated that the coupling between the N\'{e}el vector and the magnetization leads to significantly faster dynamical processes than in FMs. In the limit of low Gilbert damping, this causes strong oscillations in the N\'{e}el vector direction during the reversal process and an exchange enhancement of the switching rate compared to Brown's formula applicable to FMs. The accuracy of the analytical formula is confirmed by spin dynamics simulations. By analyzing the effect of different material parameters on the switching rate, the advantages and disadvantages of AFMs over FMs are discussed for various applications. 
Our findings contribute to the understanding of thermal effects in AFM nanostructures, their stability as well as switchability, where the latter is often affected by heating effects due to applied currents or laser excitation.

The paper is organized as follows. The analytical formulae for the switching times in uniaxial FM and AFM nanoparticles are discussed in Sec.~\ref{sec_reversal}. The spin dynamics simulations are introduced in Sec.~\ref{sec_spindyn}. The necessary conditions for the application of the macrospin model to the results of atomistic simulations are detailed in Sec.~\ref{sec_theorsim}. The switching times between FMs and AFMs are compared in Sec.~\ref{sec_comparison}, and the results are summarized in Sec.~\ref{sec_conclusion}.

\section{Analytical model}\label{sec_reversal}

\subsection{Axially symmetric FM nanoparticle}\label{sec1a}

\begin{figure}
\centering
\includegraphics[width=0.6\columnwidth]{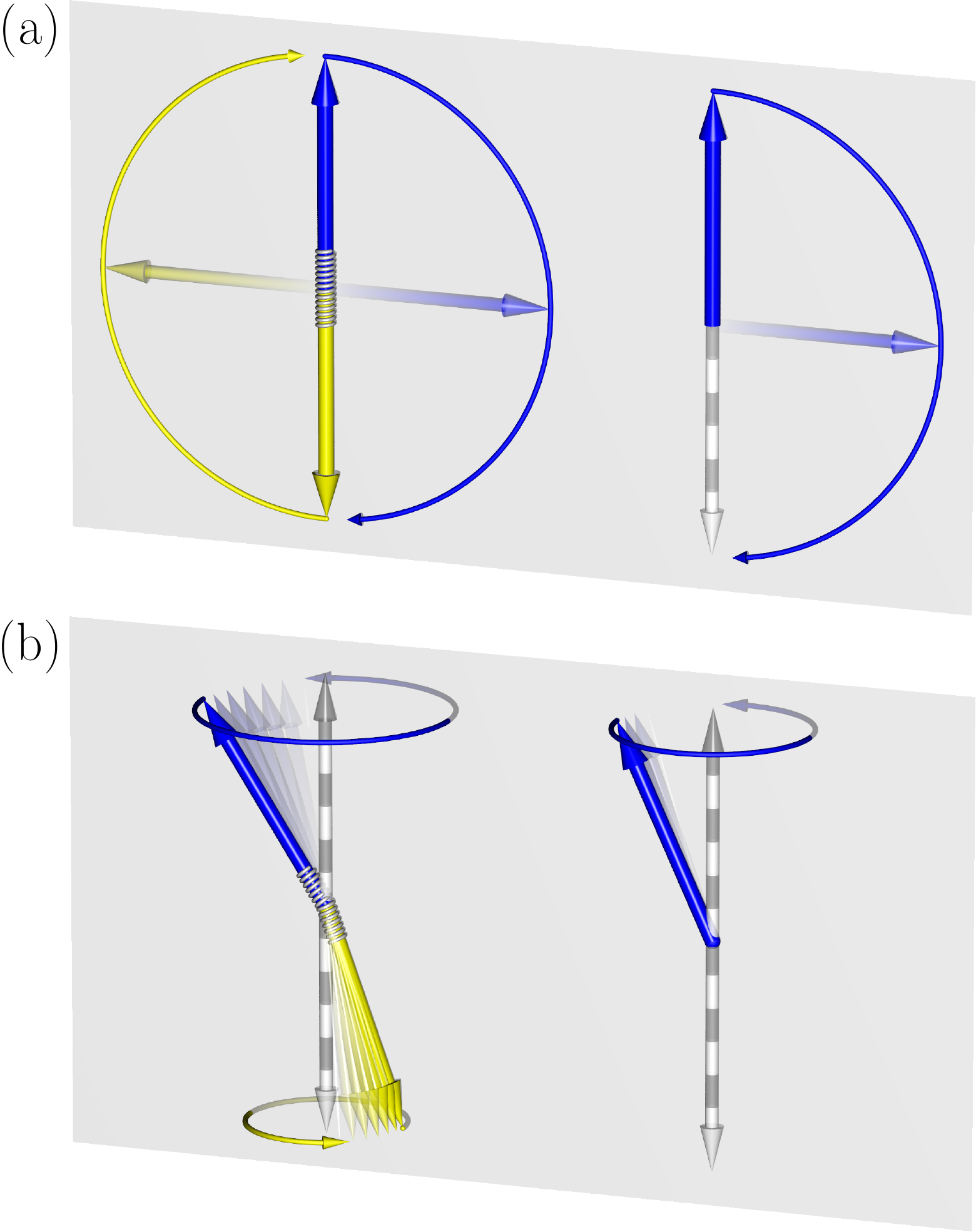}
\\
\caption{Comparison of reversal mechanisms in AFM (left) and FM (right) nanoparticles. While the energy barriers $\Delta E$ for coherent rotation are identical (a), the attempt frequencies strongly differ caused by the different dynamical properties of the eigenmodes (b). Springs in the AFM case represent that energy may be transferred between anisotropy and exchange contributions, while the latter is not present in the macrospin description of FMs.}
\label{f:sketch}
\end{figure}

For the analytical investigations we will focus on the simplest example of a magnetic nanoparticle, which switches by coherent rotation between two stable magnetic states separated by an energy barrier $\Delta E$, as sketched in Fig.~\ref{f:sketch}(a). 
We will rely on the so-called single-domain approximation, where the total magnetization of the nanoparticle is described by a single magnetic moment or macrospin in the FM case. This remains valid if the particle size stays below the exchange length, corresponding to the characteristic size of domain walls in the system.  
The dynamics can be calculated within the framework of the macroscopic Landau--Lifshitz--Gilbert equation \cite{Landau1935,Gilbert2004}
\begin{eqnarray}  
           {\dot{\mathbf{m}}} =
              - \mathbf{m}  \times  \left( \gamma\mathbf{h}_m -\alpha \frac{\dot{\mathbf{m}}}{m_{0}}\right),
             \label{eq:macro-LLG}
\end{eqnarray}
where $\alpha$ is the Gilbert damping constant, $\gamma$ is the gyromagnetic ratio, $\mathbf{m}$ the magnetization of the nanoparticle, $m_{0}$ the magnitude of the magnetization, and $\mathbf{h}_m=-\delta_\mathbf{m} \mathcal{F}$ the effective field, where $\mathcal{F}$ is the magnetic free energy of the system. For simplicity, in this work we restrain the discussion to a uniaxial particle, with the free-energy density $f=-H_{\textrm{a}} m_z^2/\left(2m_{0}\right)$, where $H_{\textrm{a}}=2\mathcal{D}_{z}N/(Vm_{0})$ is the anisotropy field, $\mathcal{D}_{z}$ is the anisotropy energy of a single spin, $N$ is the number of spins in the nanoparticle and $V$ is its volume. Equation~\eqref{eq:macro-LLG} describes the rotational motion of the macrospin, with its length fixed at $\left|\mathbf{m}\right|=m_{0}$. In this case, the free energy has two minima, $m_z/m_{0}=1$ and $m_z/m_{0}=-1$, with the energy barrier between them being $\Delta E= H_{\textrm{a}}m_{0}V/2 =\mathcal{D}_{z} N$.

Thermal activation allows the nanoparticle to jump between the free energy minima with a characteristic time scale. In the limit of low temperatures, $k_{\mathrm{B}} T \ll \Delta E$, the switching 
time for coherent rotation over the barrier was derived by Brown \cite{Brown},
\begin{equation}
 \tau_{\rm{fm}} = \frac{1+\alpha^2}{\alpha} \omega_{\textrm{a}}^{-1} \sqrt{\frac{\pi k_{\mathrm{B}} T}{\mathcal{D}_{z} N}} \textrm{e}^{\frac{\mathcal{D}_{z}N}{k_{\mathrm{B}} T}}.
 \label{eq:Ahrenius-law}
\end{equation}

This expression is of the form of the exponential N\'{e}el--Arrhenius law $\tau=\tau_{0} \textrm{e}^{\Delta E /k_{\mathrm{B}} T}$, with the energy barrier $\Delta E=\mathcal{D}_{z} N$ determined above. The prefactor $\tau_{0}$ is called the inverse attempt frequency. 
Its first factor is related to the damping dependence of the switching time, clearly with a minimum at $\alpha=1$. The second factor is the precessional time scale of the system, with $\omega_{\textrm{a}}=\gamma H_{\textrm{a}}=\gamma2\mathcal{D}_{z}N/(Vm_{0}) $. The weak temperature dependence of the prefactor is attributed to the Goldstone mode of the system at the top of the energy barrier in the axially symmetric free-energy expression.

\subsection{Nonaxially symmetric FM nanoparticles}\label{sec1b}

Equation~(\ref{eq:Ahrenius-law}) is valid for all values of the damping parameter $\alpha$. As pointed out in, e.g., Ref.~\cite{KalmykovJAP2012}, its simple form can be attributed to the fact that the Fokker--Planck equation derived from the stochastic Landau--Lifshitz--Gilbert equation simplifies to an ordinary differential equation for the polar angle variable $\cos\vartheta=m_z/m_{0}$. If the rotational symmetry of the system is broken, for example, by a tilted external magnetic field \cite{Nowak}, then the free energy $\mathcal{F}$ must describe the coupling between polar $\vartheta$ and azimuthal $\varphi$ variables, or longitudinal and transversal degrees of freedom. This transforms the Fokker--Planck equation into a partial differential equation which is significantly more difficult to handle.

For massive particles, escape rates from an energy minimum were first systematically derived by Kramers \cite{Kramers}, who differentiated between intermediate-to-high damping (IHD) and very-low-damping (VLD) limits. For IHD, it can be assumed that the system is in thermal equilibrium both close to the energy minimum ($\textrm{min}$) and in the vicinity of the saddle point ($\textrm{sp}$) which has to be crossed during the escape. The IHD limit of nonaxially symmetric FM nanoparticles was derived by Brown \cite{Brown2}, which was later revealed to be \cite{Braun,Coffey,Kachkachi,Duff,LangevinBook,Kalmykov2010} a special case of Langer's \cite{Langer} expression for multiple degrees of freedom. Within this description, the Hamiltonian or the free energy is approximated by a harmonic expansion around the minimum and close to the saddle point, while the equations of motion are linearized near the saddle point. The energy scale of thermal fluctuations is required to be much lower than the energy barrier protecting the metastable state, leading to an Arrhenius-like formula. Applications to magnetic systems can be found in, e.g., Refs.~\cite{Braun,Coffey,Dejardin,Kachkachi,Duff}. The generalization to an arbitrary number of Goldstone modes as presented in Eq.~\eqref{eqn1} below is based on harmonic transition-state theory \cite{Bessarab}, which differs from Langer's theory in applying a dynamical prefactor independent of the damping.

The switching time $\tau^{\textrm{IHD}}$ may be expressed by the formula
\begin{align}
\tau^{\textrm{IHD}}=\frac{2\pi}{\lambda_{+,\textrm{sp}}}\frac{V_{\textrm{min}}}{V_{\textrm{sp}}}\left(2\pi k_{\textrm{B}}T\right)^{\frac{P_{\textrm{sp}}-P_{\textrm{min}}}{2}}\sqrt{\frac{\prod_{j}'\left|\varepsilon_{j,\textrm{sp}}\right|}{\prod_{j}'\varepsilon_{j,\textrm{min}}}}\textrm{e}^{\frac{E_{\textrm{sp}}-E_{\textrm{min}}}{k_{\textrm{B}}T}},\label{eqn1}
\end{align}
where $E$ is the energy of the given configuration and $\varepsilon_{j}$ denotes the eigenvalues of the harmonic Hamiltonian in the equilibrium state. Ideally, all eigenvalues in the minimum are positive, and there is a single negative eigenvalue (hence the absolute value) in the first-order saddle point, along which direction the transition takes place. However, the system may possess zero-energy Goldstone modes which are to be handled separately. These must be left out of the eigenvalue products, hence the prime notation. Each of these will contribute a $\sqrt{2\pi k_{\textrm{B}}T}$ factor instead, with $P$ denoting the number of Goldstone modes. $V$ is the phase space volume belonging to the Goldstone modes. Finally, $\lambda_{+,\textrm{sp}}$ is the single positive eigenvalue of the linearized equations of motion in the saddle point. This determines how fast the system crosses the transition state. The derivation of Eq.~\eqref{eq:Ahrenius-law} based on Eq.~\eqref{eqn1} is given in Appendix~\ref{secA1a}.

However, in the VLD limit the approximations of Langer's theory break down, since the weak coupling between the system and the heat bath encapsulated in the damping parameter is no longer sufficient for ensuring thermal equilibrium at higher energy values. In order to achieve agreement with the fluctuation--dissipation theorem, one has to calculate the energy dissipated during a single precession along the energy contour including the saddle point, and ensure that this is low compared to the thermal energy $k_{\textrm{B}}T$ in the VLD case. Such a calculation for FM nanoparticles was carried out by Klik and Gunther \cite{Klik}. Finally, the missing connection between the VLD and IHD limits, the solution of the so-called Kramers turnover problem, was derived by Mel'nikov and Meshkov \cite{Melnikov} for massive particles, and adapted to nonaxially symmetric FM nanoparticles by Coffey et al. \cite{Coffey,Dejardin}. This can be summarized in the formula
\begin{eqnarray}
\tau=A^{-1}\left(\frac{\alpha S}{k_{\textrm{B}}T}\right)\tau^{\textrm{IHD}},\label{eqn0}
\end{eqnarray}
where $A\left(\dfrac{\alpha S}{k_{\textrm{B}}T}\right)$ is the depopulation factor,
\begin{eqnarray}
A\left(x\right)=\textrm{e}^{\frac{1}{2\pi}\bigintsss_{-\infty}^{\infty}\ln\left(1-\textrm{e}^{-x\left(\frac{1}{4}+y^{2}\right)}\right)\frac{1}{\frac{1}{4}+y^{2}}\textrm{d}y},\label{eqn2}
\end{eqnarray}
and $\tau^{\textrm{IHD}}$ is the switching time in the IHD limit given by Eq.~\eqref{eqn1}. The validity of the general formula for FMs was later thoroughly confirmed by the numerical solution of the Fokker--Planck equation, spin dynamics simulations and experiments; see, e.g., Refs.~\cite{Ouari2010,Kalmykov2010,KalmykovJAP2012}.

\subsection{Axially symmetric AFM nanoparticle}\label{sec1c}

For AFMs, to the best of our knowledge, analytical formulae similar to Eq.~\eqref{eq:Ahrenius-law} remain unknown. Only a few recent works have addressed the problem \cite{Ouari2010,Raikher2008}. However, they assumed AFM nanoparticles with uncompensated magnetic moments, attributed to finite-size effects and lattice defects in naturally occurring nanoparticles \cite{Neel1,Neel2}. In this limit, the AFM was effectively described as a FM with a very small magnetic moment. In spintronics applications, it is possible to prepare completely compensated AFM structures, for example by atom manipulation as demonstrated in Ref.~\cite{LothScience2012}. The dynamics in AFMs are described by coupled equations of motion for the N\'{e}el vector and the magnetization \cite{GomonayLoktev2008,HalsPRL2011,LiuPRMat2017,Yuan2018}, expected to lead to a qualitatively different behavior.
Dissipative dynamics in two-sublattice AFMs may be derived by considering two coupled Landau--Lifshitz--Gilbert equations \eqref{eq:macro-LLG} for the sublattice magnetizations $\mathbf{m}_1$ and $\mathbf{m}_2$. These are transformed to the dynamical variables of the magnetization $\mathbf{m}=(\mathbf{m}_1+\mathbf{m}_2)/2$ and the N\'eel vector $\mathbf{n}=(\mathbf{m}_1-\mathbf{m}_2)/2$. At low temperature, it is reasonable to assume that the N\'{e}el vector conserves its length $\left|\mathbf{n}\right|=m_{0}$ and only undergoes rotational motion. The magnetization remains perpendicular to the N\'{e}el vector, $\mathbf{n}\cdot\mathbf{m}=0$, since in a compensated AFM a finite magnetization may only be formed by canting the two sublattice magnetizations perpendicularly to their original 
antiparallel orientation. This leads to the equations of motion \cite{HalsPRL2011}
\begin{eqnarray}  
             \dot{\mathbf{n}} &=& - \mathbf{n}  \times  \left( \gamma\mathbf{h}_{m} - \alpha  \frac{\dot{\mathbf{m}}}{m_{0}} \right),\label{eq:macro-LLG-afm-n}   \\
           \dot{\mathbf{m}} &=&  - \mathbf{m}  \times  \left(\gamma \mathbf{h}_{m} -\alpha \frac{\dot{\mathbf{m}}}{m_{0}}  \right) - \mathbf{n}  \times  \left(  \gamma  \mathbf{h}_{n} - \alpha \frac{\dot{\mathbf{n}}}{m_{0}} \right),\:\:\:  
           \label{eq:macro-LLG-afm-m} 
\end{eqnarray}
where $\mathbf{h}_{m,n}=-\delta_{\mathbf{m},\mathbf{n}} \mathcal{F}$ are the effective fields acting on the magnetization and the N\'{e}el vector, respectively. The free-energy density of an axially symmetric single-domain AFM particle reads $f=H_{\textrm{e}}\mathbf{m}^2/\left(2m_{0}\right)-H_{\textrm{a}} n_z^2/\left(2m_{0}\right)$, with $H_{\textrm{e}}=q\mathcal{J}N/(Vm_{0})$ being the exchange field describing the coupling between the sublattices, where $q$ is the number of nearest neighbors and $\mathcal{J}$ the exchange constant in the corresponding atomistic model. In the following, we will assume that $q\mathcal{J}\gg \mathcal{D}_{z}$, which is true for practically all magnetic materials.

Although the AFM nanoparticle is still axially symmetric, it fundamentally differs from its FM counterpart described in Sec.~\ref{sec1a}. As illustrated in Fig.~\ref{f:sketch}(b), in AFMs the anisotropy energy assigned to the $z$ component of the N\'{e}el vector $n_{z}$ may transform into the exchange energy between the sublattices, leading to a finite magnetization $\mathbf{m}$, even in the conservative case. In comparison, the FM particle may only perform a precession around the easy axis with a constant polar angle $\vartheta$. Consequently, one has to rely on the theory for coupled degrees of freedom, such as in the case of nonaxially symmetric FM systems in Sec.~\ref{sec1b}, when deriving the switching time in AFMs. Applying Eq.~\eqref{eqn0} to this problem leads to the expression
\begin{eqnarray}
\tau_{\rm{afm}} = A^{-1}\left(\frac{\alpha S}{k_{\textrm{B}}T}\right)\frac{1+\alpha^2}{\alpha} \omega_{\textrm{afm}}^{-1} \sqrt{\frac{\pi k_{\mathrm{B}} T}{\mathcal{D}_{z} N}} \textrm{e}^{\frac{\mathcal{D}_{z}N}{k_{\mathrm{B}} T}},\:\:\:
 \label{eq:Ahrenius-law-afm}
\end{eqnarray}
with the derivation given in Appendix~\ref{secA1b}.

In comparison with Eq.~\eqref{eq:Ahrenius-law}, one can observe that the energy barrier $\Delta E=\mathcal{D}_{z} N$ between the minima at $n_z/m_{0}=1$ and $n_z/m_{0}=-1$ remains the same in 
Eq.~\eqref{eq:Ahrenius-law-afm}, as long as all individual spins rotate coherently during switching. Similarly, the temperature-dependent square-root term attributed to the axial symmetry is preserved. On the other hand, the frequency $\omega_{\textrm{a}}$ is replaced by $\omega_{\textrm{afm}}$,
\begin{align}
&\omega_{\textrm{afm}}=\nonumber
\\
&\frac{\gamma N}{Vm_{0}}\left[\left(\mathcal{D}_{z}-\frac{1}{2}q\mathcal{J}\right)+\sqrt{\left(\frac{1}{2}q\mathcal{J}+\mathcal{D}_{z}\right)^{2}+\frac{2\mathcal{D}_{z}q\mathcal{J}}{\alpha^2}}\right].\:\:\:\:\:\: \label{eq:frequency}
\end{align}

The transition between the IHD and the VLD limits is governed by the ratio of the thermal energy $k_{\textrm{B}}T$ and the energy loss per cycle on the contour including the saddle point,
\begin{eqnarray}
\alpha S=\alpha N\left(\frac{16\mathcal{D}_{z}^{2}}{3\sqrt{2\mathcal{D}_{z}q\mathcal{J}}}+4\sqrt{2\mathcal{D}_{z}q\mathcal{J}}\right);\label{eqn13}
\end{eqnarray}
for the derivation see Appendix~\ref{secA1c}.

In order to highlight the differences and similarities between the FM and AFM cases, appropriate asymptotic expressions are derived. On the one hand, in the limit of high damping $\alpha\gg 1$, one has $\omega_{\textrm{afm}}\approx\omega_{\textrm{a}}$ and $A\left(\alpha S/\left(k_{\textrm{B}}T\right)\right)\approx 1$ due to the strong energy dissipation $\alpha S\gg k_{\textrm{B}}T$, leading to
\begin{eqnarray}
\tau_{\rm{afm},\alpha\gg 1} \approx \alpha \omega_{\textrm{a}}^{-1} \sqrt{\frac{\pi k_{\mathrm{B}} T}{\mathcal{D}_{z} N}} \textrm{e}^{\frac{\mathcal{D}_{z}N}{k_{\mathrm{B}} T}},\:\:\:
 \label{eq:Langerhigha}
\end{eqnarray}
which coincides with the asymptotic behavior for FMs, Eq.~\eqref{eq:Ahrenius-law}.

On the other hand, significant deviations may be observed between the two types of systems in the limit of low damping. For $\alpha\ll 1$, the characteristic frequency of AFMs is $\alpha\omega_{\textrm{afm}}\approx\sqrt{2\mathcal{D}_{z}q\mathcal{J}}=\sqrt{q\mathcal{J}/\left(2\mathcal{D}_{z}\right)}\omega_{\textrm{a}}$, indicating that the dynamics are exchange-enhanced compared to FMs. Furthermore, the depopulation factor may be approximated as $A\left(\alpha S/\left(k_{\textrm{B}}T\right)\right)\approx \alpha S/\left(k_{\textrm{B}}T\right)\approx \alpha 4N\sqrt{2\mathcal{D}_{z}q\mathcal{J}}/\left(k_{\textrm{B}}T\right)$ for slow energy dissipation and $q\mathcal{J}\gg \mathcal{D}_{z}$. The VLD limit of Eq.~\eqref{eq:Ahrenius-law-afm} reads
\begin{eqnarray}
\tau_{\rm{afm},\alpha\ll 1} \approx \frac{1}{\alpha}\frac{k_{\textrm{B}}T}{4q\mathcal{J}N} \omega_{\textrm{a}}^{-1} \sqrt{\frac{\pi k_{\mathrm{B}} T}{\mathcal{D}_{z} N}} \textrm{e}^{\frac{\mathcal{D}_{z}N}{k_{\mathrm{B}} T}}.\:\:\:
 \label{eq:Langerlowa}
\end{eqnarray}

In Eq.~\eqref{eq:Langerlowa}, the switching time is inversely proportional to the damping parameter, as expected from the fluctuation--dissipation theorem \cite{Kramers}. Furthermore, it is reduced by a factor of $k_{\textrm{B}}T/\left(4q\mathcal{J}N\right)$ compared to the appropriate limit of Eq.~\eqref{eq:Ahrenius-law}. The typical value of intrinsic damping in magnetic materials is $\alpha=0.001-0.01$, e.g., $\alpha=0.0025$ was determined for Mn$_2$Au in Ref.~\cite{Bhattacharjee2018}. This means that the switching time in AFMs could be up to several orders of magnitude shorter than in FMs, which effectively means much less thermal stability.

The high- and low-damping limits of the AFM switching time, defined by Eqs.~\eqref{eq:Langerhigha} and \eqref{eq:Langerlowa}, may be connected by the simplified formula
\begin{eqnarray}
\tau_{\rm{afm,asymptotic}}=\frac{\frac{k_{\textrm{B}}T}{4q\mathcal{J}N}+\alpha^{2}}{\alpha}\omega_{\textrm{a}}^{-1}\sqrt{\frac{\pi k_{\textrm{B}}T}{\mathcal{D}_{z} N}}\textrm{e}^{\frac{\mathcal{D}_{z} N}{k_{\textrm{B}}T}},\label{eqn17}
\end{eqnarray}
which has an analogous form to Eq.~\eqref{eq:Ahrenius-law}. This clearly expresses the difference in the behavior between FMs and AFMs: while for the former the minimal switching time is found for $\alpha_{\textrm{fm,min}}=1$, for the latter this value now depends on the system parameters, $\alpha_{\textrm{afm,min}}=\sqrt{k_{\textrm{B}}T/\left(4q\mathcal{J}N\right)}$, being decreased due to the exchange interaction. Since for high $\alpha$ values the switching times in FMs and AFMs coincide, while the minimum is shifted to lower $\alpha$ values in AFMs, this implies that AFM nanoparticles are significantly less resistant against thermal fluctuations at low damping compared to their FM counterparts. However, note that in the immediate vicinity of $\alpha_{\textrm{afm,min}}$, Eq.~\eqref{eq:Ahrenius-law-afm} is expected to give a more accurate description than Eq.~\eqref{eqn17}, since the former includes a more precise interpolation between the VLD and IHD limits exactly in this turnover regime.

\section{Spin dynamics simulations}\label{sec_spindyn}

To test the validity of Eqs.~\eqref{eq:Ahrenius-law} and \eqref{eq:Ahrenius-law-afm}, we performed atomistic spin dynamics simulations. 
For the description of the magnetic system, we introduce the classical atomistic spin Hamiltonian
\begin{eqnarray}
\mathcal{H}=\mp\frac{1}{2}\sum_{\left<i,j\right>}J\mathbf{S}_{i}\mathbf{S}_{j}
-\sum_{i}D_{z}S_{i,z}^{2}.
\label{eqn1sd}
\end{eqnarray}
Here the $\mathbf{S}_{i}$ variables denote unit vectors on a simple cubic lattice and $J$ is the Heisenberg exchange interaction between atoms at nearest-neighbor sites $i$ and $j$. 
For the $-$ sign in Eq.~\eqref{eqn1sd} the ground state is FM, while for the $+$ sign it is AFM. $D_{z}>0$ is the single-ion magnetocrystalline anisotropy, implying that the ground state of the system lies along the $z$ direction.

The time evolution of the unit vectors $\mathbf{S}_i$ is described by the Landau--Lifshitz--Gilbert equation, 
\begin{eqnarray}  
           (1+\alpha^2)\mu_{\textrm{s}}{\dot{\mathbf{S}}}_i =
              - \gamma  \mathbf{S}_i  \times  \left[ \mathbf{H}_i 
             +  \alpha   \left( \mathbf{S}_i  \times \mathbf{H}_i  \right) \right],
             \label{eq:LLG}
\end{eqnarray}
where $\mu_{\textrm{s}}$ denotes the magnetic moment of a single spin and $\alpha$ is the Gilbert damping as in the macrospin model.
By including a Langevin thermostat, the 
equilibrium and nonequilibrium thermodynamic properties can be obtained in the classical approximation. 
The effective local magnetic field at lattice site $i$ is  
\begin{equation} 
\mathbf{H}_i= - \frac{\partial \mathcal{H}}{\partial \mathbf{S}_i} + \boldsymbol{\xi}_i(t),
\end{equation} 
where $\mathcal{H}$ is given by Eq.~\eqref{eqn1sd} in the present case and
$\boldsymbol{\xi}_i$ is a field-like stochastic process.
Here we consider the white-noise limit \cite{NowakBook}, with the first and second moments
\begin{equation}
\langle \boldsymbol{\xi}_i(t) \rangle = \boldsymbol{0}, \quad 
 \langle \xi_{i,a}(0) \xi_{j,b}(t) \rangle = 
 \frac{2 \alpha k_{\mathrm{B}}T \mu_{\textrm{s}}}{\gamma} \delta_{ij}\delta_{ab} \delta(t), 
 \label{eq:noise-correlator}
\end{equation}
where $a$ and $b$ denote the Cartesian components.

\section{Correspondence between theory and simulations}\label{sec_theorsim}

\subsection{Temperature-dependent effective parameters}\label{sec2a}

For a direct comparison of Eqs.~\eqref{eq:Ahrenius-law} and \eqref{eq:Ahrenius-law-afm} with the results of the spin dynamics simulations, it has to be ensured that the assumptions which the analytical formulae are based on are satisfied by the atomistic model. As long as the linear size of the system is shorter than a characteristic length scale on the order of the exchange length, $L_{\textrm{e}} \sim \sqrt{J/D_z}$, it is expected that coherent rotation is the primary mechanism of magnetization reversal in the nanoparticles. 
Above this threshold, the nucleation of a pair of domain walls becomes energetically favorable compared to the energy barrier which has to be overcome by coherent rotation \cite{hinzkeJMMM00,BraunPRL, braunBOOK00, hinzkePRB00, hinzkePRB98}.

Even for small particles, one has to take into account that in the atomistic model the thermal fluctuations decrease $m_{0}$, the equilibrium length of the magnetization in FMs and of the N\'{e}el vector in AFMs \cite{Nowak2005}. In earlier publications for FM systems \cite{EvansPRB2015}, it was found that the dimensionless magnetization may be well approximated by the phenomenological relation $m_{0}V/(N\mu_{\textrm{s}})=(1-T/T_c)^{1/3}$ for 3d Heisenberg models.
Furthermore, one has to account for finite-size effects. 
Small systems such as the nanoparticles considered here have a reduced magnetization compared to the bulk at a given temperature, as a result of lower coordination numbers at the surfaces. 
For 3d Heisenberg spin models, finite-size-scaling theory provides a value for the apparent Curie temperature as a function of the size $L$ (linear characteristic size of the nanoparticle), $T_c(L)/T_c^{\infty}= 1-\left(d_0/L\right)^{1/\nu}$, where
the parameter $d_0$ corresponds to the characteristic exchange length, and $\nu$ to the critical exponent. A recent work \cite{EllisAPL2015} in FM FePt nanoparticles using similar parameters to our simulations has estimated $d_0=0.4\,\textrm{nm}$, to be compared to a lattice constant of $a=0.38\,\textrm{nm}$, and $\nu=0.856$. The critical temperature of the cubic Heisenberg model in the thermodynamic limit was found to be $k_{\textrm{B}}T_c^{\infty}=1.443$ $J$ \cite{Peczak}. In this work we performed simulations for a cubic nanoparticle composed of $N=4^{3}=64$ spins; therefore, the lateral size is 4 spins, meaning $d_{0}/L=0.4/(0.38\times 4)=0.238$ in the finite-size-scaling expression, leading to $T_c(L)=1.173$ $J$. We found that the phenomenological relation using this critical temperature was in agreement with spin dynamics simulations of the dimensionless order parameter at the temperature ranges where coherent reversal is dominant.

In the analytical expressions the effect of the reduced order parameter may be considered by assuming temperature-dependent magnetic parameters in Eqs.~\eqref{eq:Ahrenius-law} and \eqref{eq:Ahrenius-law-afm} \cite{Nowak2005},
\begin{eqnarray}
\mathcal{D}_{z}&=&D_{z}\left(\frac{m_{0}V}{N\mu_{\textrm{s}}}\right)^{3},\label{eq:dzTdep}
\\
\mathcal{J}&=&J\left(\frac{m_{0}V}{N\mu_{\textrm{s}}}\right)^{2}.\label{eq:JTdep}
\end{eqnarray}
The cubic dependence of the anisotropy on the dimensionless magnetization expressed in Eq.~\eqref{eq:dzTdep} is the result of the Callen--Callen theory \cite{Callen,Callen2}. The quadratic dependence of the exchange in Eq.~\eqref{eq:JTdep} may be derived from the random phase approximation \cite{Tyablikov}.

Furthermore, the reduced coordination number $q$ at the surface also directly affects Eq.~\eqref{eq:Ahrenius-law-afm}. Here we substituted the mean value of the number of nearest neighbors: for a nanoparticle composed of $N=64$ spins in simple cubic arrangement, $q=6$ for the spins inside $(2^3=8)$, $q=5$ for the spins at the faces $(6\times 2\times 2=24)$, $q=4$ for the spin at the edges $(12\times 2=24)$, and $q=3$ for the spins at the corners $(8)$, thus $q_{\rm{avg}}=4.5$.

\subsection{Oscillations in the order parameter in the VLD limit of AFM nanoparticles}\label{sec2b}

A further requirement for an accurate comparison between simulations and analytical expressions is that the identified switching events in the simulations have to correspond to the reversals described by the theory \cite{Kalmykov2010}. For uniaxial nanoparticles with easy axis along the $z$ direction, the following criteria may be identified. First, the $z$ component of the order parameter $\mathbf{m}$ or $\mathbf{n}$ has to change sign, and thereafter cross a threshold value governed by the equilibrium value $m_{0}$ at the given temperature. During the process, the energy of the particle increases while crossing the energy barrier, before decreasing again when coming to rest in the other energy minimum; see Supplemental Videos 1 and 2 \cite{supp} for an illustration of this process.

\begin{figure}
	\centering
	\includegraphics[width=\columnwidth]{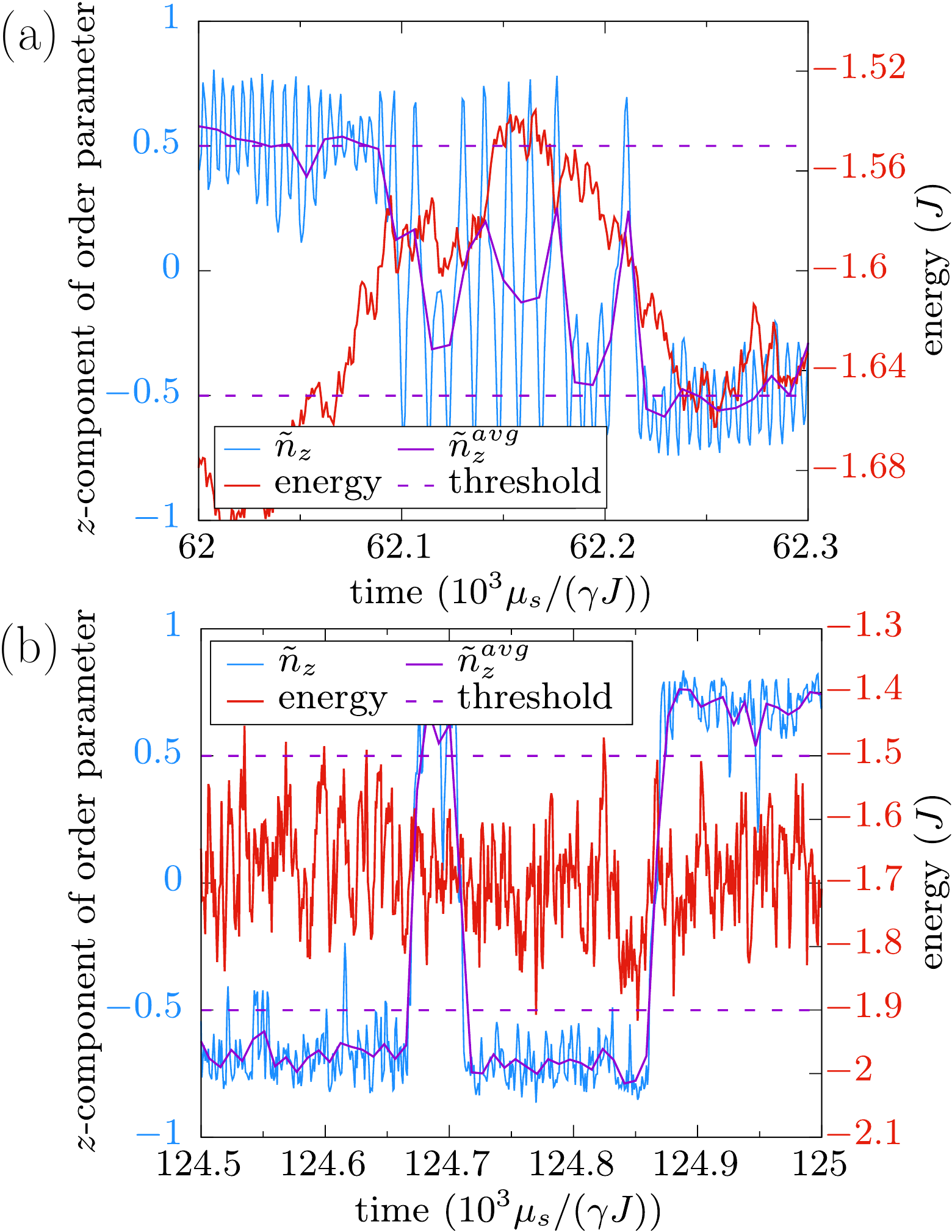}
	\\
	\caption{
		Illustration of the switching events in the AFM nanoparticle for (a) low ($\alpha=0.0005$) and (b) intermediate ($\alpha=0.1$) damping. The other simulation parameters are $T=0.6 \ J/k_\mathrm{B}$, $D_z=0.1\ J$, for a cubic nanoparticle consisting of $N=4^3=64$ spins. The threshold values for the switching are chosen to be $\pm0.75\left<\left|\tilde{n}_{z}\right|\right>$, where $\tilde{n}_{z}$ is the $z$ component of the dimensionless order parameter and $\left<\left|\tilde{n}_{z}\right|\right>$ is the thermal average of its absolute value. $\tilde{n}_{z}^{\textrm{avg}}$ was obtained by performing a moving average on the $\tilde{n}_{z}$ data using a window of width $\Delta t=8.8$ $\mu_{\textrm{s}}/\left(\gamma J\right)$.}	\label{fig:oscillations}
\end{figure}

For FMs, the sign change of $m_{z}$ is always accompanied by an increase in the anisotropy energy. On the other hand, in AFM nanoparticles the energy can be transformed between the anisotropy contribution of the N\'{e}el vector and the exchange contribution of the magnetization, meaning that $n_{z}$ may switch sign even if the total energy of the system remains constant. 
In the low-damping limit such an oscillatory motion can indeed be observed, where the $z$ component of the N\'{e}el vector switches sign and crosses the threshold value many times before coming to rest in one of the minima, see Fig.~\ref{fig:oscillations}(a) and Supplemental Video 3 \cite{supp}. This is analogous to a mechanical particle in a double-well potential, where the energy is transformed between the kinetic and potential parts during the motion. During these oscillations in the N\'{e}el vector, the energy of the system is slowly varied due to the weak coupling to the heat bath, meaning that the oscillations take place on a roughly constant energy surface and hence they only represent a single switching event. For an estimate of the oscillation periods see Appendix~\ref{secA1d}.

To determine the actual switching events in the low-damping limit in the simulations, we therefore used a time average of the data, where the time window was larger than the period of the fast oscillations of the N\'{e}el vector while crossing the energy barrier. As shown in Fig.~\ref{fig:oscillations}(a), in the averaged data the $z$ component of the dimensionless order parameter only crosses the threshold value once after its sign change during a single reversal. 
In contrast, for intermediate-to-high values of $\alpha$ the energy fluctuates strongly on the time scale of a single rotation, and the oscillatory switching is absent as shown in Fig.~\ref{fig:oscillations}(b). In this case, the same number of switching events are registered both with and without the averaging procedure.

\section{Comparison of switching times}\label{sec_comparison}

\begin{figure}
\centering
\includegraphics[width=\columnwidth]{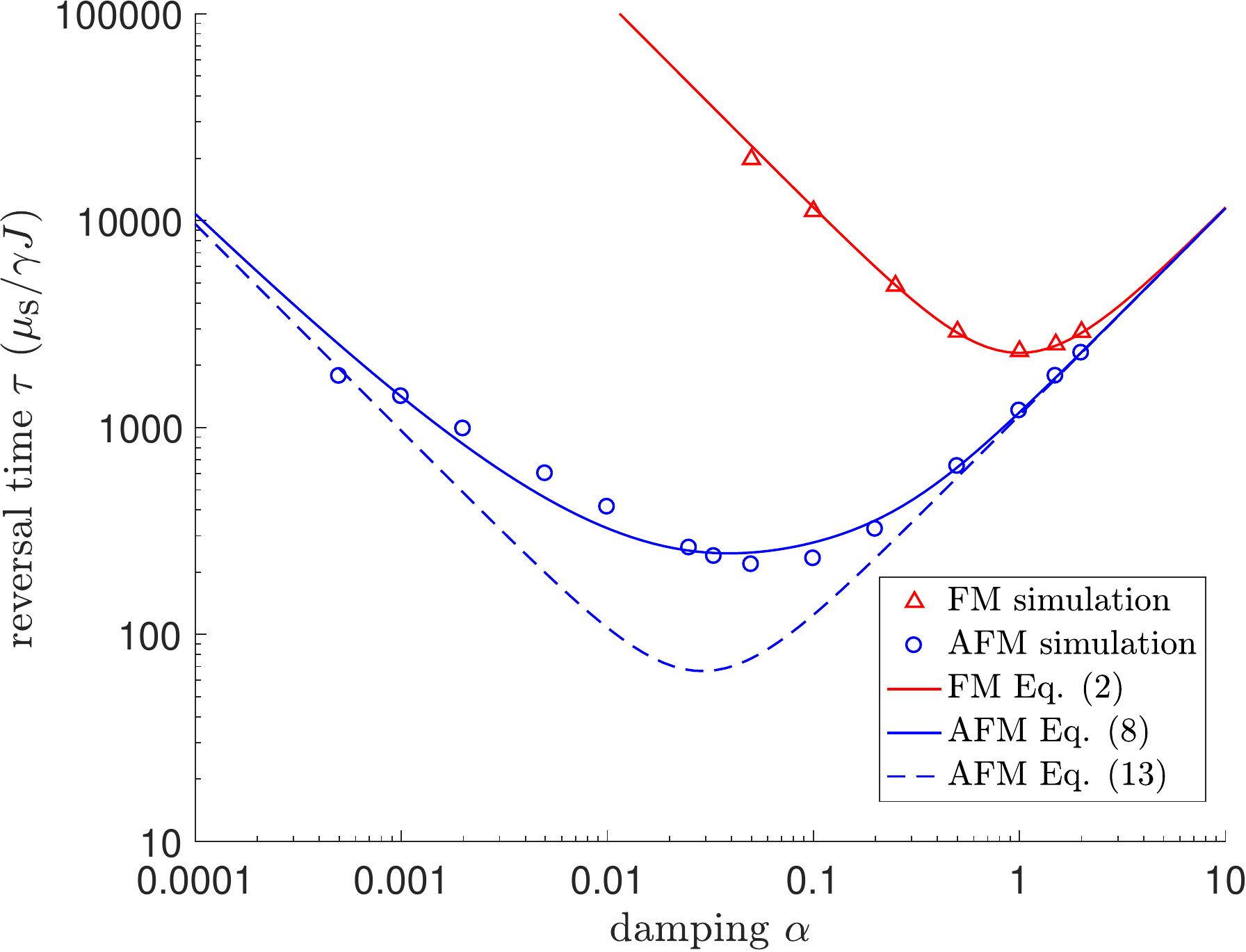}
\\
\caption{
Damping dependence of the switching time for FM and AFM nanoparticles. The system parameters are $T=0.6 \ J/k_\mathrm{B}$, $D_{z}=0.1\ J$, $N=64$. Symbols correspond to simulations using atomistic spin dynamics methods and lines to the analytical formulae Eqs.~\eqref{eq:Ahrenius-law}, \eqref{eq:Ahrenius-law-afm}, and \eqref{eqn17}.
}
\label{fig:Fig_3_paper}
\end{figure}

In order to validate the damping dependence of the switching time in both FMs and AFMs, we performed computer simulations by varying the damping value $\alpha$ at a fixed temperature $T=0.6 \ J/k_\mathrm{B}$, shown in Fig.~\ref{fig:Fig_3_paper}. In order to enable an accurate comparison, the same absolute value of the exchange interaction $J$ and the anisotropy $D_z=0.1\ J$ was considered during the simulations, performed for a cubic nanoparticle consisting of $N=64$ spins. As can be seen in the figure, Eq.~\eqref{eq:Ahrenius-law} gives good agreement with the simulation results for the FM case, while Eq.~\eqref{eq:Ahrenius-law-afm} is accurate for the AFM case over the whole parameter range. While the switching times are similar for high damping, the minimal switching time is found for significantly lower $\alpha$ values in the AFM case, leading to a reduced thermal stability in the limit of low damping. Note that the asymptotic expression Eq.~\eqref{eqn17} for AFMs, which has an analogous form to Eq.~\eqref{eq:Ahrenius-law} for FMs, underestimates the switching time in the turnover regime. In particular, for the present simulation parameters the VLD limit, characterized by the relation $\tau_{\rm{afm},\alpha\ll 1}\propto\alpha^{-1}$ in Eq.~\eqref{eq:Langerlowa}, is not reached yet for $\alpha\approx 0.001$, and the AFM switching time shows a weaker dependence on the damping in this turnover regime.

\begin{figure}
\centering
\includegraphics[width=\columnwidth]{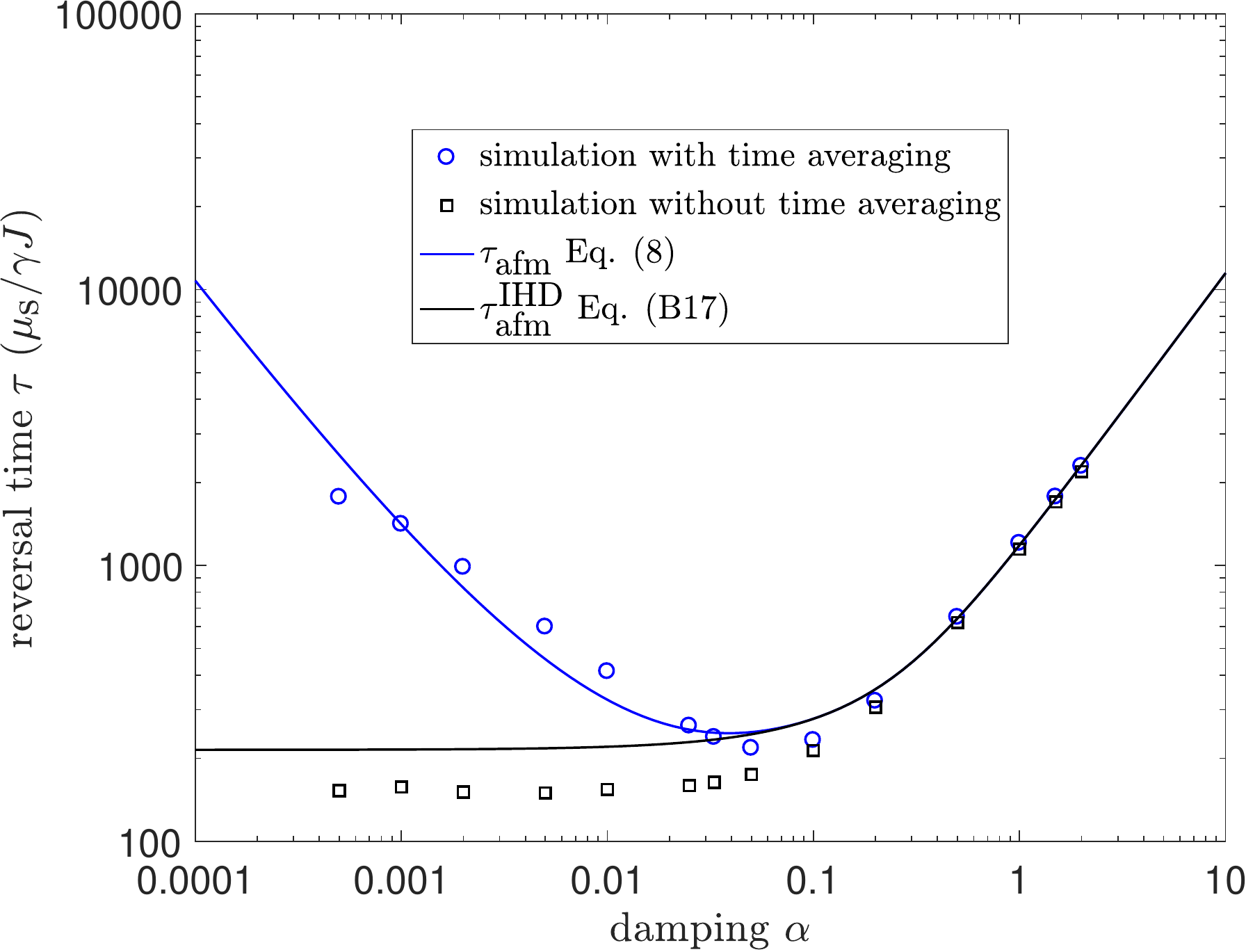}
\\
\caption{
Damping dependence of the switching time for the antiferromagnetic nanoparticle, using the parameters $T=0.6 \ J/k_\mathrm{B}$, $D_z=0.1\ J$, $N=64$. Circles and squares correspond to the same simulation data as in Fig.~\ref{fig:Fig_3_paper}, with and without performing the time averaging. Lines show Eq.~\eqref{eq:Ahrenius-law-afm}, expected to hold for all $\alpha$ values, and Eq.~\eqref{eq:Ahrenius-law-afmA} without the depopulation factor, which is only applicable in the intermediate-to-high-damping limit.
}
\label{fig:Fig_3_SI}
\end{figure}

Figure~\ref{fig:Fig_3_SI} illustrates the effect of time averaging of the simulation data on the obtained switching times. Moving averages were performed on a time interval of $\Delta t=8.8$ $\mu_{\textrm{s}}/\left(\gamma J\right)$.
Without performing the time averaging, the mean time between sign changes of the $z$ component of the order parameter converges to a constant value at low damping, similarly to the intermediate-to-high-damping formula, given by Eq.~\eqref{eq:Ahrenius-law-afmA} in Appendix~\ref{secA1b}. However, this behavior is in contradiction with the fluctuation--dissipation theorem. The range in $\alpha$ where the time-averaging starts to play a significant role in the simulation data coincides with the interval where the depopulation factor in Eq.~\eqref{eq:Ahrenius-law-afm} becomes important in the theoretical description. This emphasizes the necessity of correctly determining the switching time in the very-low-damping-limit both in the analytical model as well as in the numerical simulations.

\begin{figure}
\centering
\includegraphics[width=\columnwidth]{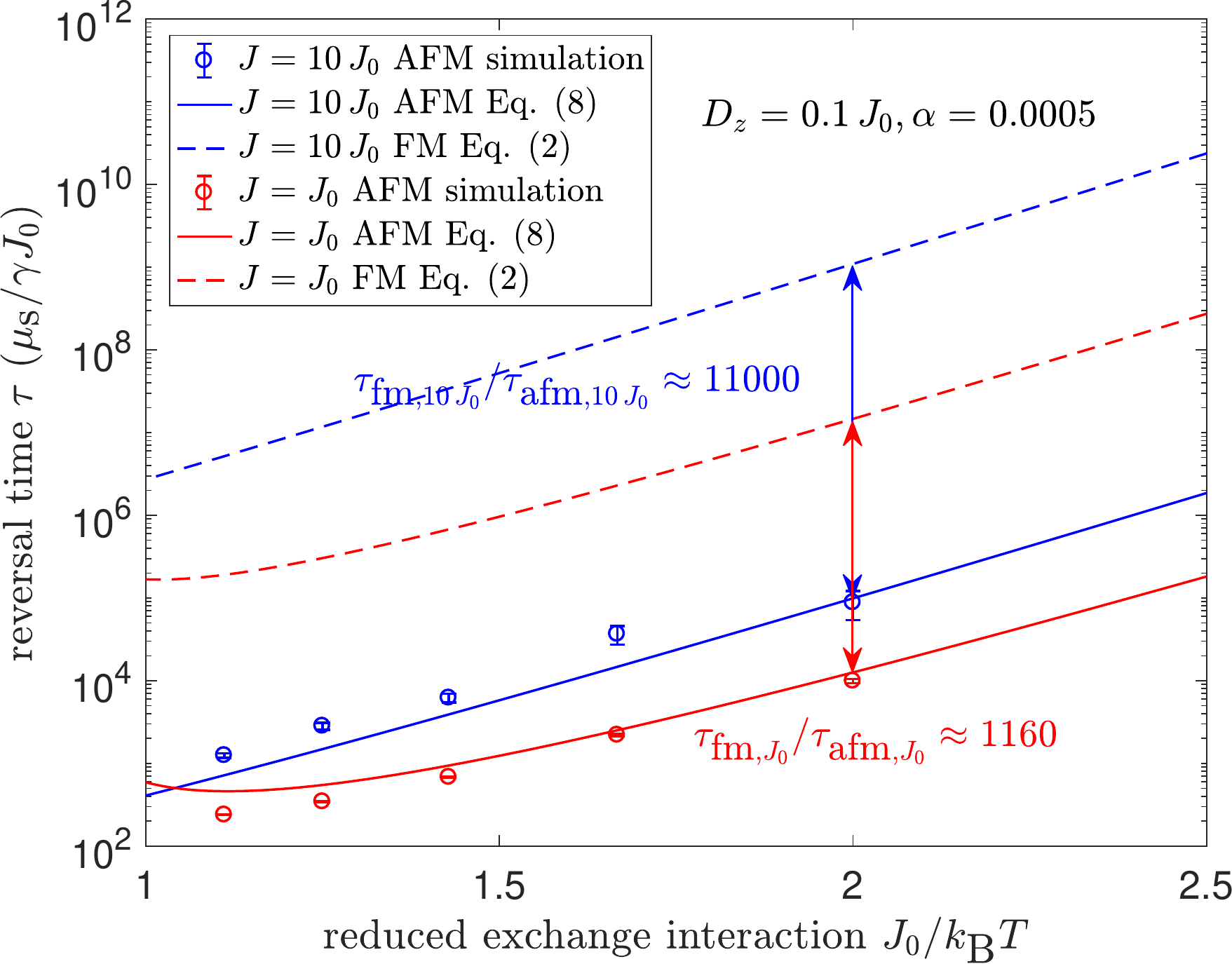}
\\
\caption{
Dependence of the switching time on the exchange interaction $J$ for FM and AFM nanoparticles. The system parameters are $\alpha=0.0005$, $D_{z}=0.1\ J_{0}$, $N=64$. Symbols correspond to simulations using atomistic spin dynamics methods for the AFM case and lines to the analytical formulae
Eqs.~\eqref{eq:Ahrenius-law} and \eqref{eq:Ahrenius-law-afm}.
}
\label{fig:Fig_4_paper}
\end{figure}

A further important difference between FMs and AFMs, as can be deduced from Eqs.~\eqref{eq:Ahrenius-law} and \eqref{eq:Ahrenius-law-afm}, is that the switching time in AFMs depends on the microscopic exchange interaction $\mathcal{J}$, while this parameter is absent in the single-domain description of FMs. The analytical expressions Eqs.~\eqref{eq:Ahrenius-law} and \eqref{eq:Ahrenius-law-afm} for different values of $J$ are compared in Fig.~\ref{fig:Fig_4_paper} as a function of temperature, using the parameters $D_z=0.1\ J_{0}$ and $\alpha=0.0005$. For the AFM case simulation results are also presented, confirming the assumed N\'{e}el--Arrhenius law in this parameter range. 
For the FM case with the significantly longer switching times only the analytical formula Eq.~\eqref{eq:Ahrenius-law} is shown, which had been confirmed in earlier publications \cite{KalmykovJAP2012} and in Fig.~\ref{fig:Fig_3_paper} here for a different damping regime. Note that while Eq.~\eqref{eq:Ahrenius-law} does not explicitly depend on $\mathcal{J}$, the predicted analytical curves are still different for $J=J_{0}$ and $J=10$ $J_{0}$, since the equilibrium magnetization $m_{0}$ is higher in the latter case (cf. Eqs.~\eqref{eq:dzTdep} and \eqref{eq:JTdep}). As indicated in the figure, at the lowest temperature where the simulations were performed ($J_{0}/k_\mathrm{B}T=2$), the ratio $\tau_{\rm{fm}}/\tau_{\rm{afm}}$ is about $10$ times larger for $10$ times higher exchange interaction, in agreement with the VLD damping limits of Eqs.~\eqref{eq:Ahrenius-law} and \eqref{eq:Langerlowa}.

\section{Conclusion}\label{sec_conclusion}

In summary, we investigated the superparamagnetic limit of AFM nanoparticles analytically as well as by means of computer simulations. The derived analytical expression, Eq.~\eqref{eq:Ahrenius-law-afm}, for the mean switching time indicates a drastically reduced thermal stability of AFM nanostructures as compared to their FM counterparts because of the exchange enhancement of the attempt frequency. The latter is caused by the coupling between the anisotropy term connected to the N\'{e}el vector and the exchange term connected to the magnetization in the free-energy density of single-domain AFMs, which also causes a strong oscillation of the N\'{e}el vector direction at low damping values during the switching process.

The significantly faster dynamics in AFMs is one of their main proposed advantages over FMs in spintronics applications \cite{OlejnikScienceAdvance18}. However, this enhanced speed also leads to an increased susceptibility to thermal fluctuations as demonstrated here; for realistic materials with a low damping value, the switching times of AFMs can be expected to be four to five orders of magnitude shorter than those of FMs, a finding that is in agreement with a work on antiferromagnetic grains in exchange bias systems \cite{FernandezAPL2010}. Furthermore, the procedures capable of increasing the switching times in FMs may be less efficient in AFMs. The energy barrier in the Arrhenius expressions Eqs.~\eqref{eq:Ahrenius-law} and \eqref{eq:Ahrenius-law-afm}, which is the leading term in the temperature dependence, may be increased by choosing a higher anisotropy value $\mathcal{D}_{z}$, a larger system size $N$, a lower temperature $T$, or at larger order parameters $m_{0}$ achieved by coupling the microscopic spins stronger to each other by a higher exchange coupling $\mathcal{J}$. According to the VLD limit Eq.~\eqref{eq:Langerlowa}, all of these methods except increasing the anisotropy lead to a decrease in the inverse attempt frequency, meaning that they decrease the $\tau_{\rm{afm}}/\tau_{\rm{fm}}$ ratio assuming the same system parameters. Furthermore, for FMs damping values in the range $\alpha=0.001-0.01$, typical for materials suggested for spintronic devices \cite{Bhattacharjee2018}, will surely fall into the VLD regime, where lower $\alpha$ values lead to an enhanced switching time. On the other hand, for AFMs similar values may belong to the turnover region where the lifetime is minimal and the dependence on $\alpha$ is weak, around $\alpha_{\textrm{afm,min}}\approx\sqrt{k_{\textrm{B}}T/\left(4q\mathcal{J}N\right)}$. These problems may be circumvented by selecting materials with a high damping value, where the difference between FM and AFM switching times disappears. 

However, fast reversal of the nanoparticles may also be desired in specific applications. Since thermal activation facilitates the current-induced switching in spintronic devices \cite{Meinert}, a higher attempt frequency necessitates a lower current density for achieving the same switching rate. In magnetic hyperthermia \cite{Pankhurst},  the reversal of nanoparticles is used to provide targeted warming of tissues, which can become more efficient at higher frequencies. For such purposes, AFM nanoparticles may provide advantages over their FM counterparts.

\begin{acknowledgments}
Financial support for this work at the University of Konstanz was provided by the Deutsche Forschungsgemeinschaft via SFB 1214. At the FU Berlin support by the Deutsche Forschungsgemeinschaft through SFB/TRR 227  "Ultrafast Spin Dynamics", Project A08 is gratefully acknowledged. L.R. would like to acknowledge the Alexander von Humboldt Foundation and the National Research, Development and Innovation Office of Hungary via Project No. K115575 for support.
\end{acknowledgments}

\appendix

\section{FM switching time in the IHD limit\label{secA1a}}

Here the switching time of axially symmetric FM nanoparticles, given by Eq.~\eqref{eq:Ahrenius-law}, is derived based on the general expression Eq.~\eqref{eqn1}. The free-energy density is given by $f=-H_{\textrm{e}} m_z^2/\left(2m_{0}\right)$ where $H_{\textrm{e}}=2\mathcal{D}_{z}N/(Vm_{0})$, and the normalization $\left|\mathbf{m}\right|=m_{0}$ is assumed.
The expression has a minimum at $m_{z}/m_{0}=1$, and the expansion is performed in the small variables $m_{x}/m_{0},m_{y}/m_{0}\ll 1$. This yields
\begin{eqnarray}
\mathcal{F}_{\textrm{min}}&=&-\mathcal{D}_{z}N,\label{eqn1a}
\\
\varepsilon_{1,\textrm{min}}&=&\varepsilon_{2,\textrm{min}}=2\mathcal{D}_{z}N.\label{eqn1b}
\end{eqnarray}

The saddle point is at $m_{x}/m_{0}=1$ with the expansion variables $m_{y}/m_{0},m_{z}/m_{0}\ll 1$, which results in
\begin{eqnarray}
\mathcal{F}_{\textrm{sp}}&=&0,\label{eqn1c}
\\
\varepsilon_{1,\textrm{sp}}&=&-2\mathcal{D}_{z}N,\label{eqn1d}
\\
\varepsilon_{2,\textrm{sp}}&=&0.\label{eqn1e}
\end{eqnarray}

Note that $\varepsilon_{1,\textrm{sp}}$ is negative, corresponding to the unstable mode in the saddle point. The other eigenvalue $\varepsilon_{2,\textrm{sp}}$ describes a Goldstone mode, representing the fact that the saddle point can be arbitrarily chosen along the circle $m_{x}^{2}+m_{y}^{2}=m_{0}^{2}$. The corresponding phase space volume is
\begin{eqnarray}
V_{\textrm{sp}}=2\pi,\label{eqn1f}
\end{eqnarray}
the circumference of the circle.

The linearized Landau--Lifshitz--Gilbert equation in the saddle point reads
\begin{eqnarray}
\partial_{t}m_{y}&=&\frac{1}{1+\alpha^{2}}\frac{\gamma N}{Vm_{0}}2\mathcal{D}_{z}m_{z}=\frac{1}{1+\alpha^{2}}\omega_{\textrm{a}} m_{z},\label{eqn9}
\\
\partial_{t}m_{z}&=&\frac{1}{1+\alpha^{2}}\frac{\gamma N}{Vm_{0}}\alpha 2\mathcal{D}_{z}m_{z}=\frac{\alpha}{1+\alpha^{2}} \omega_{\textrm{a}} m_{z},\label{eqn8}
\end{eqnarray}
with the eigenvalues
\begin{eqnarray}
\lambda_{1,\textrm{sp}}&=&\frac{\alpha}{1+\alpha^{2}} \omega_{\textrm{a}},\label{eqn9a}
\\
\lambda_{2,\textrm{sp}}&=&0,\label{eqn9b}
\end{eqnarray}
where $\lambda_{+,\textrm{sp}}=\lambda_{1,\textrm{sp}}$ is the single positive eigenvalue.

Substituting Eqs.~(\ref{eqn1a})-(\ref{eqn1f}) and Eq.~(\ref{eqn9a}) into Eq.~(\ref{eqn1}) gives precisely Eq.~(\ref{eq:Ahrenius-law}), which in this special case is valid for all values of the damping.

\section{AFM switching time in the IHD limit\label{secA1b}}

Here Eq.~\eqref{eq:Ahrenius-law-afm} without the depopulation factor will be derived based on Eq.~\eqref{eqn1}. We will use the free-energy density $f=H_{\textrm{e}}\mathbf{m}^2/\left(2m_{0}\right)-H_{\textrm{a}} n_z^2/\left(2m_{0}\right)$, with $H_{\textrm{e}}=q\mathcal{J}N/(Vm_{0})$ being the exchange field, where $q$ is the number of nearest neighbors and $\mathcal{J}$ the exchange constant in the corresponding atomistic spin model, rescaled by accounting for the thermally reduced order parameter.

The minimum of the free energy $\mathcal{F}$ is at $\mathbf{n}/m_{0}=\left(0,0,1\right),\mathbf{m}/m_{0}=\left(0,0,0\right)$ with
\begin{eqnarray}
\mathcal{F}_{\textrm{min}}&=&-\mathcal{D}_{z}N,\label{eqn44}
\\
\varepsilon_{1,\textrm{min}}&=&\varepsilon_{2,\textrm{min}}=2\mathcal{D}_{z}N,\label{eqn45}
\\
\varepsilon_{3,\textrm{min}}&=&\varepsilon_{4,\textrm{min}}=q\mathcal{J}N.\label{eqn46}
\end{eqnarray}

For $\mathcal{D}_{z}\ll q\mathcal{J}$ the saddle point is $\mathbf{n}/m_{0}=\left(1,0,0\right),\mathbf{m}/m_{0}=\left(0,0,0\right)$, where the expansion yields
\begin{eqnarray}
\mathcal{F}_{\textrm{sp}}&=&0,\label{eqn47}
\\
\varepsilon_{1,\textrm{sp}}&=&-2\mathcal{D}_{z}N,\label{eqn48}
\\
\varepsilon_{2,\textrm{sp}}&=&0,\label{eqn49}
\\
\varepsilon_{3,\textrm{sp}}&=&\varepsilon_{4,\textrm{sp}}=q\mathcal{J}N,\label{eqn50}
\end{eqnarray}

Here $\varepsilon_{1,\textrm{sp}}$ is the unstable mode and $\varepsilon_{2,\textrm{sp}}$ is the Goldstone mode with
\begin{eqnarray}
V_{\textrm{sp}}=2\pi.\label{eqn50a}
\end{eqnarray}

The linearized equations of motion in the saddle point read
\begin{eqnarray}
\partial_{t}m_{y}&=&\frac{\gamma N}{Vm_{0}}2\mathcal{D}_{z}n_{z}-\alpha\partial_{t}n_{z},\label{eqn51}
\\
\partial_{t}m_{z}&=&\alpha\partial_{t}n_{y},\label{eqn52}
\\
\partial_{t}n_{y}&=&-\frac{\gamma N}{Vm_{0}}q\mathcal{J}m_{z}-\alpha\partial_{t}m_{z},\label{eqn53}
\\
\partial_{t}n_{z}&=&\frac{\gamma N}{Vm_{0}}q\mathcal{J}m_{y}+\alpha\partial_{t}m_{y},\label{eqn54}
\end{eqnarray}
leading to the eigenvalues
\begin{align}
\lambda_{1,\textrm{sp}}=&0,\label{eqn55}
\\
\lambda_{2,\textrm{sp}}=&-\frac{1}{1+\alpha^{2}}\frac{\gamma N}{Vm_{0}}\alpha q\mathcal{J},\label{eqn56}
\\
\lambda_{3,\textrm{sp}}=&\frac{1}{1+\alpha^{2}}\frac{\gamma N}{Vm_{0}}\Bigg[\alpha \left(\mathcal{D}_{z}-\frac{1}{2}q\mathcal{J}\right)\Bigg.\nonumber
\\
&\Bigg.+\sqrt{\alpha^{2}\left(\frac{1}{2}q\mathcal{J}+\mathcal{D}_{z}\right)^{2}+2\mathcal{D}_{z}q\mathcal{J}}\Bigg],\label{eqn57}
\\
\lambda_{4,\textrm{sp}}=&-\frac{1}{1+\alpha^{2}}\frac{\gamma N}{Vm_{0}}\Bigg[\alpha \left(\frac{1}{2}q\mathcal{J}-\mathcal{D}_{z}\right)\Bigg.\nonumber
\\
&\Bigg.+\sqrt{\alpha^{2}\left(\frac{1}{2}q\mathcal{J}+\mathcal{D}_{z}\right)^{2}+2\mathcal{D}_{z}q\mathcal{J}}\Bigg],\label{eqn58}
\end{align}
where the positive eigenvalue is $\lambda_{+,\textrm{sp}}=\lambda_{3,\textrm{sp}}$.

Substituting Eqs.~(\ref{eqn44})-(\ref{eqn50a}) and Eq.~(\ref{eqn57}) into Eq.~(\ref{eqn1}) produces
\begin{align}
\tau_{\rm{afm}}^{\textrm{IHD}} =& \frac{1+\alpha^2}{\alpha} \frac{Vm_{0}}{\gamma N}\left[\left(\mathcal{D}_{z}-\frac{1}{2}q\mathcal{J}\right)\right.\nonumber
\\
&\left.+\sqrt{\left(\frac{1}{2}q\mathcal{J}+\mathcal{D}_{z}\right)^{2}+\frac{2\mathcal{D}_{z}q\mathcal{J}}{\alpha^2}}\right]^{-1} \sqrt{\frac{\pi k_{\mathrm{B}} T}{\mathcal{D}_{z} N}} \textrm{e}^{\frac{\mathcal{D}_{z}N}{k_{\mathrm{B}} T}},\:\:\:
 \label{eq:Ahrenius-law-afmA}
\end{align}
the intermediate-to-high-damping limit of Eqs.~\eqref{eq:Ahrenius-law-afm} and \eqref{eq:frequency}. Note that since the eigenvalues $\varepsilon_{3,\textrm{min}},\varepsilon_{4,\textrm{min}}$ cancel with $\varepsilon_{3,\textrm{sp}},\varepsilon_{4,\textrm{sp}}$, the difference between the ferromagnetic and antiferromagnetic cases only comes from the dynamical prefactor $\lambda_{+,\textrm{sp}}$, which is exchange-enhanced at low and intermediate damping for the latter.

\section{Energy dissipation per cycle when passing through the saddle point\label{secA1c}}

Here the depopulation factor in Eq.~\eqref{eqn0} will be calculated for the AFM particle. The variable $S$ in the argument of $A$ in Eq.~(\ref{eq:Ahrenius-law-afm}) denotes the action of the undamped motion crossing through the saddle point. Equation~(\ref{eqn0}) expresses that if $\alpha S$, the energy dissipated during a single cycle of motion over the saddle point \cite{Dejardin,Klik}, is small compared to the thermal energy $k_{\textrm{B}}T$, then it takes longer for the particle to cross the energy barrier since it can no longer be assumed that the equilibrium Maxwell--Boltzmann distribution is formed in the region close to the saddle point.

In order to calculate this energy dissipation, Eqs.~\eqref{eq:macro-LLG-afm-n} and \eqref{eq:macro-LLG-afm-m} are linearized in $\alpha$ at low damping, yielding
\begin{eqnarray}  
             \dot{\mathbf{n}} &=& - \gamma\mathbf{n}  \times  \left( \mathbf{h}_{m} + \alpha \frac{\mathbf{m}}{m_{0}} \times\mathbf{h}_{m}+ \alpha \frac{\mathbf{n}}{m_{0}}\times\mathbf{h}_{n}\right),\label{eqn5}   \\
           \dot{\mathbf{m}} &=&  - \gamma\mathbf{m}  \times  \left( \mathbf{h}_{m} + \alpha \frac{\mathbf{m}}{m_{0}} \times\mathbf{h}_{m}+ \alpha \frac{\mathbf{n}}{m_{0}} \times\mathbf{h}_{n}  \right) \nonumber
           \\
           &&- \gamma\mathbf{n}  \times  \left(    \mathbf{h}_{n} + \alpha \frac{\mathbf{n}}{m_{0}}\times\mathbf{h}_{m} \right).  
           \label{eqn6} 
\end{eqnarray}

The free energy dissipation per cycle may be written as
\begin{eqnarray}
-\Delta \mathcal{F}&=&\alpha S=-\int_{0}^{T}\dot{\mathcal{F}}\textrm{d}t=\int_{0}^{T}\int\mathbf{h}_{m}\dot{\mathbf{m}}+\mathbf{h}_{n}\dot{\mathbf{n}}\textrm{d}\boldsymbol{r}\textrm{d}t\nonumber
\\
&=&\alpha\gamma m_{0}V\int_{0}^{T}\left(\frac{\mathbf{m}}{m_{0}} \times\mathbf{h}_{m}+\frac{\mathbf{n}}{m_{0}} \times\mathbf{h}_{n}\right)^{2}\nonumber
\\
&&+\left(\frac{\mathbf{n}}{m_{0}} \times\mathbf{h}_{m}\right)^{2}\textrm{d}t.\label{eqn7}
\end{eqnarray}

Introducing the renormalized variables $\hat{\mathbf{m}}=\mathbf{m}/m_{0}$, $\hat{\mathbf{n}}=\mathbf{n}/m_{0}$, and substituting $\mathbf{h}_{m}=-q\mathcal{J}N/(Vm_{0})\hat{\mathbf{m}}, \mathbf{h}_{n}=2\mathcal{D}_{z}N/(Vm_{0})\hat{n}_{z}\mathbf{e}_{z}$ for the considered system one obtains
\begin{eqnarray}
\alpha S=\frac{\alpha\gamma N^{2}}{Vm_{0}}\int_{0}^{T}4\mathcal{D}_{z}^{2}\left(1-\hat{n}_{z}^{2}\right)\hat{n}_{z}^{2}+\left(q\mathcal{J}\right)^{2}\hat{\mathbf{m}}^{2}\textrm{d}t,\:\label{eqn8new}
\end{eqnarray}
with the integral to be evaluated along the trajectory of the undamped motion crossing the saddle point.

For $\alpha=0$, Eqs.~\eqref{eqn5} and \eqref{eqn6} may be written as
\begin{eqnarray}
\partial_{t}\hat{m}_{x}&=&-\frac{\gamma N}{Vm_{0}}2\mathcal{D}_{z}\hat{n}_{y}\hat{n}_{z},\label{eqn66}
\\
\partial_{t}\hat{m}_{y}&=&\frac{\gamma N}{Vm_{0}}2\mathcal{D}_{z}\hat{n}_{x}\hat{n}_{z},\label{eqn67}
\\
\partial_{t}\hat{m}_{z}&=&0,\label{eqn68}
\\
\partial_{t}\hat{n}_{x}&=&\frac{\gamma N}{Vm_{0}}q\mathcal{J}\left(\hat{n}_{y}\hat{m}_{z}-\hat{n}_{z}\hat{m}_{y}\right),\label{eqn69}
\\
\partial_{t}\hat{n}_{y}&=&\frac{\gamma N}{Vm_{0}}q\mathcal{J}\left(\hat{n}_{z}\hat{m}_{x}-\hat{n}_{x}\hat{m}_{z}\right),\label{eqn70}
\\
\partial_{t}\hat{n}_{z}&=&\frac{\gamma N}{Vm_{0}}q\mathcal{J}\left(\hat{n}_{x}\hat{m}_{y}-\hat{n}_{y}\hat{m}_{x}\right),\label{eqn71}
\end{eqnarray}
for the axially symmetric AFM nanoparticle. Since the constraint $\left|\hat{\mathbf{n}}\right|=1$ is satisfied by the dynamical equations, the normalized N\'{e}el vector may be rewritten in spherical coordinates, $\left(\hat{n}_{x},\hat{n}_{y},\hat{n}_{z}\right)=\left(\sin\vartheta\cos\varphi,\sin\vartheta\sin\varphi,\cos\vartheta\right)$. For the variable $\hat{\mathbf{m}}$ one has $\hat{\mathbf{n}}\cdot\hat{\mathbf{m}}=0$, and its $z$ component is a constant of motion as expressed by Eq.~\eqref{eqn68}. Without the damping, the free energy of the system is also conserved during the time evolution,
\begin{eqnarray}
\mathcal{F}=\frac{q\mathcal{J}}{2}N\hat{\mathbf{m}}^{2}-\mathcal{D}_{z}N\cos^{2}\vartheta.\label{eqn74}
\end{eqnarray}

Using the conserved quantities $\mathcal{F}$ and $\hat{m}_{z}$, Eqs.~\eqref{eqn66}-\eqref{eqn71} may be expressed as
\begin{eqnarray}
\partial_{t}\vartheta&=&\mp\sqrt{\omega^{2}_{\mathcal{F}}-\omega^{2}_{0}\sin^{2}\vartheta-\frac{\omega^{2}_{\hat{m}_{z}}}{\sin^{2}\vartheta}},\label{eqn75}
\\
\partial_{t}\varphi&=&-\frac{1}{\sin^{2}\vartheta}\omega_{\hat{m}_{z}},\label{eqn75a}
\end{eqnarray}
with
\begin{eqnarray}
\omega_{0}&=&\frac{\gamma N}{Vm_{0}}\sqrt{2\mathcal{D}_{z}q\mathcal{J}},\label{eqn76}
\\
\omega_{\mathcal{F}}&=&\frac{\gamma N}{Vm_{0}}\sqrt{2\left(\frac{\mathcal{F}}{N}+\mathcal{D}_{z}\right)q\mathcal{J}},\label{eqn77}
\\
\omega_{\hat{m}_{z}}&=&\frac{\gamma N}{Vm_{0}}q\mathcal{J}\hat{m}_{z}.\label{eqn77a}
\end{eqnarray}

For the trajectory including the saddle point one has $\mathcal{F}=0$, see Eq.~\eqref{eqn47}, and $\hat{m}_{z}=0$. Equation~\eqref{eqn75} may be used to change the parametrization from the time $t$ to the polar angle $\vartheta$, which simplifies Eq.~\eqref{eqn8new} to
\begin{eqnarray}
\alpha S&=&\alpha N\int_{0}^{2\pi}\frac{4\mathcal{D}_{z}^{2}}{\sqrt{2\mathcal{D}_{z}q\mathcal{J}}}\left(\left|\cos\vartheta\right|-\left|\cos\vartheta\right|^{3}\right)\nonumber
\\
&&+\sqrt{2\mathcal{D}_{z}q\mathcal{J}}\left|\cos\vartheta\right|\textrm{d}\vartheta.\label{eqn12}
\end{eqnarray}

Evaluating the integral Eq.~\eqref{eqn12} yields Eq.~\eqref{eqn13}.

\section{Oscillations in the N\'{e}el vector based on the theoretical model\label{secA1d}}

\begin{figure}
	\centering
	\includegraphics[width=0.9\columnwidth]{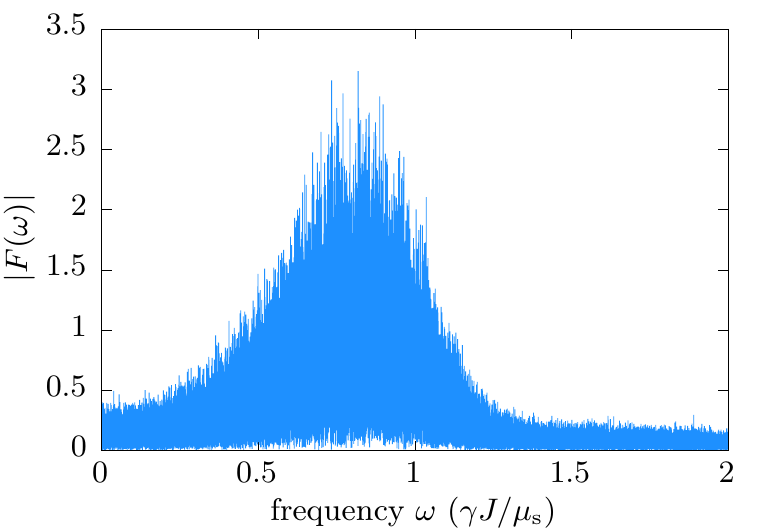}
	\\
	\caption{
		Fourier spectrum of the oscillations of the $z$ component of the order parameter from the spin dynamics simulations. The same simulation parameters were used as for Fig.~\ref{fig:oscillations}(a), $\alpha=0.0005$, $T=0.6 \ J/k_\mathrm{B}$, $D_z=0.1\ J$, $N=64$. The characteristic oscillation frequency from Eq.~\eqref{eqn76} is $\omega_{0}=0.66$ $\gamma J/\mu_{\textrm{s}}$.} 
	\label{fig:spectra_saddle}
\end{figure}

As shown in Fig.~\ref{fig:oscillations}, significant oscillations in the $z$ component of the order parameter were observed in the spin dynamics simulations of antiferromagnetic nanoparticles at very low damping values. This can be explained by the fact that for $\mathcal{F}>0$ where the switching occurs, even in the conservative system $\hat{\mathbf{n}}$ will perform full rotations during which its $z$ component changes sign, as described by Eq.~\eqref{eqn75}. For $\hat{m}_{z}=0$, the period of these oscillations may be evaluated in a closed form,
\begin{eqnarray}
T_{\mathcal{F}}=\!\!\!\bigintss_{0}^{2\pi}\!\!\!\!\!\!\!\frac{1}{\sqrt{\omega^{2}_{\mathcal{F}}-\omega^{2}_{0}\sin^{2}\vartheta}}\textrm{d}\vartheta=\frac{4}{\omega_{\mathcal{F}}}K\left(\frac{\omega_{\mathcal{F}}}{\omega_{0}}\right)\!\!,\label{eqn78}
\end{eqnarray}
with $K$ the complete elliptic integral of the first kind.

It can be seen from Eq.~(\ref{eqn78}) that the oscillation frequency will change as the free energy varies due to the coupling to the heat bath. If the thermal fluctuations are weak as required for the application of Arrhenius-like expressions such as Eqs.~\eqref{eq:Ahrenius-law} and \eqref{eq:Ahrenius-law-afm}, the free energy does not become significantly higher than its saddle-point value during the switching, and in this case the oscillation frequencies will be comparable to $\omega_{0}$. 
For example, $0.01\le \mathcal{F}/\left(\mathcal{D}_{z}N\right)\le 0.2$ yields $0.39\le 2\pi/\left(T_{\mathcal{F}}\omega_{0}\right)\le 0.65$. The adiabatic variation of the energy leads to a wide distribution of frequency values if the oscillations are investigated in Fourier space, as displayed in Fig.~\ref{fig:spectra_saddle}. Using the temperature-dependent effective parameters described in Sec.~\ref{sec2a}, for the model coefficients in Fig.~\ref{fig:spectra_saddle} one obtains $\omega_{0}=0.66$ $\gamma J/\mu_{\textrm{s}}$, roughly corresponding to the peak in the frequency distribution. The period of these characteristic oscillations is $2\pi/\omega_{0}=9.48$ $\mu_{\textrm{s}}/\left(\gamma J\right)$, meaning that the averaging window of $\Delta t=8.8$ $\mu_{\textrm{s}}/\left(\gamma J\right)$ used in Figs.~\ref{fig:oscillations} and \ref{fig:Fig_3_paper} is expected to suppress most of the oscillatory switching events, and lead to a comparable mean reversal time to the one predicted by the analytical model.

\end{document}